\documentclass[aps,prl,twocolumn,superscriptaddress,showpacs,floatfix,longbibliography]{revtex4-2}
\usepackage{graphicx,bm,epsfig,textcomp,color,dcolumn,setspace,array,mathrsfs,amsmath,amssymb,gensymb,booktabs,url,hyperref,bibentry,natbib}

\begin{document}

\title{Identifying Axion Insulator by Quantized Magnetoelectric Effect in Antiferromagnetic ${\mathrm{MnBi}}_{2}{\mathrm{Te}}_{4}$ Tunnel Junction}

\author{Yu-Hang Li}
\thanks{yuhang.li@ucr.edu}
\affiliation{Department of Electrical and Computer Engineering, University of California, Riverside, California 92521, USA}
\author{Ran Cheng}
\thanks{rancheng@ucr.edu}
\affiliation{Department of Electrical and Computer Engineering, University of California, Riverside, California 92521, USA}
\affiliation{Department of physics and Astronomy, University of California, Riverside, California 92521, USA}

\begin{abstract}
Intrinsic magnetic topological insulator ${\mathrm{MnBi}}_{2}{\mathrm{Te}}_{4}$ is believed to be an axion insulator in its antiferromagnetic ground state. However, direct identification of axion insulators remains experimentally elusive because the observed vanishing Hall resistance, while indicating the onset of the axion field, is inadequate to distinguish the system from a trivial normal insulator. Using numerical Green's functions, we theoretically demonstrate the quantized magnetoelectric current in a tunnel junction of atomically thin ${\mathrm{MnBi}}_{2}{\mathrm{Te}}_{4}$ sandwiched between two contacts, which is a smoking-gun signal that unambiguously confirms antiferromagnetic ${\mathrm{MnBi}}_{2}{\mathrm{Te}}_{4}$ to be an axion insulator. Our predictions can be verified directly by experiments.
\end{abstract}

\maketitle

Recently, topological insulators with intrinsic magnetism becomes a new frontier dubbed intrinsic magnetic topological insulators (MTIs), where the time reversal symmetry is broken by the spontaneous magnetic ordering rather than magnetic disorders~\cite{Tokura2019Magnetic,Deng2020QUantum,Li2019DIrac,Li2021Spin,Li2021Critical,Song2021Delocalization}, holding great potential for the realization of high temperature topological materials. Since the topological phases of intrinsic MTIs are highly mingled with the magnetic states, manipulating the magnetic ordering through external magnetic fields, temperature or thickness will simultaneously tune the correlated topological states~\cite{Li2019Intrinsic,Gu2021}. For example, depending on the magnetic states, ${\mathrm{MnBi}}_{2}{\mathrm{Te}}_{4}$ (MBT) can exhibit versatile topological phases such as topological insulator~\cite{Gong2019Experimental}, (high Chern number) Chern insulator~\cite{Deng2020QUantum,Ge2020high}, quantum spin Hall insulator and Weyl semimetal~\cite{Li2019Intrinsic}, and in particular, axion insulator~\cite{Liu2020Robust}.

Unlike other topological phases characterized by the first Chern number~\cite{Thouless1982Chern,Hasan2010}, an axion insulator is in a higher order topological phase characterized by the symmetry protected axion field $\theta=\pi$~\cite{Qi2008Topological,Qi2011Topological,Zhang2020Mobius,Xu2019Higher}, which can manifest as the quantized topological magnetoelectric (TME) effect~\cite{Sekine2021,Zhao2021,Nenno2020} and other striking transport phenomena~\cite{Wu2016QUantized,Tse2010Giant,Nomura2011Surface,Qi2009Inducing,Gao2021Layer}. However, because the first Chern number of an axion insulator vanishes identically, the ensuing transport effect on a Hall bar device exhibits a vanishing Hall resistance accompanied by a large longitudinal resistance, which is just similar to a normal insulator. This property makes it rather elusive to properly distinguish axion insulators from normal insulators by transport experiments~\cite{Liu2020Robust}. Therefore, to confirm the existence of axion insulator, a viable experimental scheme without ambiguity is needed.

In this Letter, we propose an axion insulator tunnel junction consisting of a few-layer MBT sandwiched between two metallic contacts as an experimental setup to unambiguously identify axion insulators through the quantized TME. We first show that a perpendicular magnetic field can induce a surface charge polarization that is physically related to the layer-resolved Chern numbers and the quantized axion field $\theta=\pi$. When the magnetic field adiabatically varies with time (\textit{i.e.}, with a frequency far less than the insulating gap), the surface charge polarization becomes time-dependent and will generate an AC charge current through the tunnel junction. We use time-dependent non-equilibrium Green's function to quantity the detectable AC current driven by a harmonic magnetic field, which agrees remarkably well with the time derivative of the induced charge polarization, thus strengthening the validity and reliability of our proposed scheme to identify axion insulators. Since archetypal materials parameters have been adopted in the modeling, we anticipate our theory to be able to inspire and guide experiments in the perceivable future.

\textit{Low energy effective Hamiltonian.---}
MBT is a van der Waals magnet consisting of Te-Bi-Te-Mn-Te-Bi-Te septuple layers (SLs) arranged on a triangle lattice with parallel intralayer ferromagnetic order while adjacent SLs are coupled antiferromagnetically. Under the basis $\begin{bmatrix}\lvert p_z^+,\uparrow\rangle,&\lvert p_z^-,\uparrow\rangle,&\lvert p_z^+,\downarrow\rangle,&\lvert p_z^-,\downarrow\rangle\end{bmatrix}^T$ with $\lvert p_z^{+(-)},\sigma\rangle$ the spin-$\sigma$ orbital of Bi (Te), the low energy Hamiltonian for a layered MBT can be written as~\cite{Zhang2019Topological,Lian2020Flat,Li2021Spin}
\begin{align}
\mathcal{H}=\sum_{a=0}^3d_a(\bm{k})\Gamma_a+\Delta\sum_i\bm{m}_i\cdot\bm{\sigma}\otimes\tau_0.
\label{eq:Hamiltonian}
\end{align} 
Here, the first term is an effective Hamiltonian for a three dimensional topological insulator, where $d_0(\bm{k})=M_0-B_1k_z^2-B_2(k_x^2+k_y^2)$, $d_1(\bm{k})=A_2k_x$, $d_2(\bm{k})=A_2k_y$, $d_3(\bm{k})=A_1k_z$ with $A_{1(2)}$, $B_{1(2)}$, $M_0$ being system parameters and the lattice momentum $\bm{k}=\begin{pmatrix}k_x,&k_y,&k_z\end{pmatrix}$. $\Gamma_{0}=\sigma_0\otimes\tau_3$ and $\Gamma_{a}=\sigma_a\otimes\tau_1$ ($a=1,2,3$) where $\sigma_a$ and $\tau_a$ are Pauli matrices acting on the spin and orbital spaces, respectively. The second term describes the exchange interaction between topological electrons and magnetic ordering, where $\Delta$ is the exchange strength and $\bm{m}_i$ is the unit magnetization vector of the $i$-th SL~\cite{Li2021Spin}. Henceforth in all numerical calculations, the materials parameters are shown in Tab.~\ref{parameters}, and temperature is set to be zero. 

\begin{table}[t]
	\centering
	\caption{Parameters adopted. $a_0$ is the lattice constant. $M_0$, $A_{1(2)}$ and $B_{1(2)}$ are based in Ref.~\cite{Zhang2019Topological}. $J$, $K$ and $M_S$ are chosen from Ref.~\cite{Yang2021Odd}. $\mu_B$ is the Bohr magneton. The exchange gap $\Delta$ is evaluated from Refs.~\cite{Otrokov2019Prediction,Zeugner2019}.
			}
	\begin{tabular}{ccccccccc}
		\hline
		\hline
		$a_0$(nm)&$\Delta$(eV) &$M_0$(eV) &$A_1$(eV$\cdot$ nm) & $A_2$(eV$\cdot$ nm) \\
		\hline
		5&-0.05&-0.1165&0.27232&0.31964\\
		\hline
		\hline
		$J$(meV) & $K$(meV) & $M_S$ ($\mu_B$)& $B_1$(eV$\cdot \text{nm}^2$)&$B_2$(eV$\cdot\text{nm}^2$) \\
		\hline
		0.68&0.21&5/2&0.119048&0.094048\\
		\hline
		\hline
	\end{tabular}
	\label{parameters}
\end{table}

Since the topological states of MBT is intertwined with the magnetic ordering, we first need to determine its magnetic configuration. In the macro-spin approximation (spatially uniform magnetization within a particular SL), the magnetic property of an $N$-SL MBT can be characterized by the free energy~\cite{Mills1968Surface,*Mills1968Sur}
\begin{align}
U=J\sum_{i=1}^{N-1}\bm{m}_i\cdot\bm{m}_{i+1}-\sum_{i=1}^{N}\left[\frac{K}{2}(m_i^z)^2+M_s\bm{\mathcal{B}}\cdot\bm{m}_i\right],
\label{Mills_model}
\end{align}
where $J$ is the antiferromagnetic interlayer exchange interaction, $K$ is the easy-axis anisotropy, $\bm{\mathcal{B}}$ is external magnetic field, and $M_s$ is the saturation magnetization of each SL. The magnetization vector is parameterized as $\bm{m}_i=\{\sin\theta_i\cos\phi_i,\sin\theta_i\sin\phi_i,\cos\theta_i\}$ with $\theta_i$ ($\phi_i$) the polar (azimuthal) angle. Without losing generality, we assume that $\bm{\mathcal{B}}$ is applied along $z$ direction and $\bm{m}_i$ rotates only in the $xz$ plane. We obtain the equilibrium magnetic configuration by minimizing the free energy $U$ using the steepest descent method~\cite{Rossler2004Magnetic,*Rossler2006Spin}, which is detailed in the supplementary materials (SM)~\cite{Supply}.

Figure.~\ref{Mag_Hall}(a) shows the total magnetization as a function of the applied magnetic field for a six-SL MBT, where we identify the spin-flop critical points at around $\mathcal{B}_c^{\pm}\approx\pm3$T, beyond which the Zeeman energy overcomes the exchange and anisotropy interactions and induces non-collinear spin configurations until the system is fully polarized into a ferromagnetic state at above $10$T (see Fig. S1 in the SI). Such distinct magnetic evolution is in quantitative agreement with experiments~\cite{Deng2020QUantum,Liu2020Robust}. The complicated spin configurations in the intermediate spin-flop phases are discussed in the SM~\cite{Supply}.

\begin{figure}[t]
  \centering
  \includegraphics[width=\linewidth]{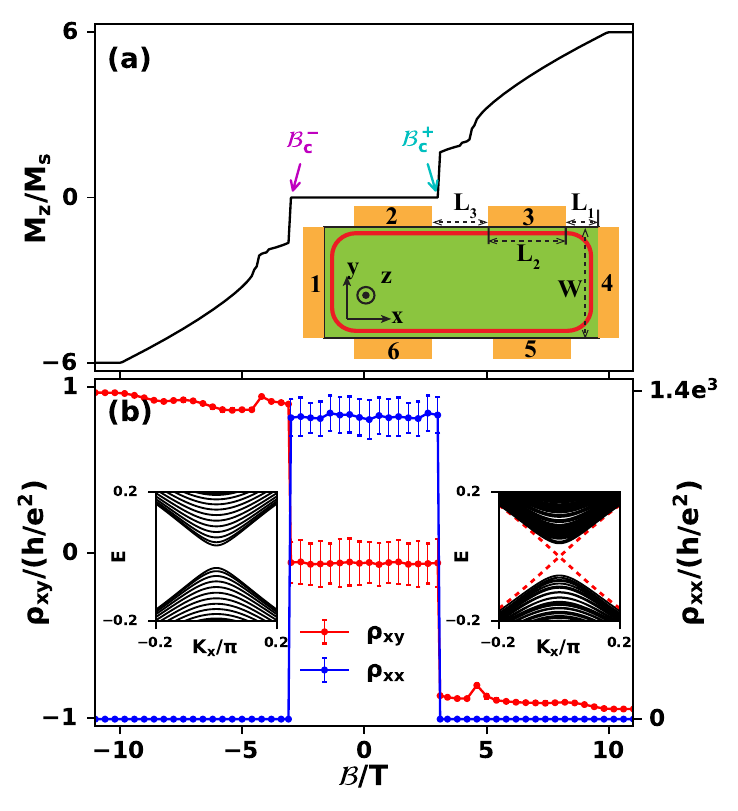}
  \caption{
	(a) Total magnetization as a function of magnetic field for a 6-SL MBT. Inset: schematics of a Hall bar device where the metallic leads are shaded orange and the MBT is colored green. The red curve marks the edge channel when the MBT becomes a Chern insulator (\textit{i.e.}, when $|\mathcal{B}|>|\mathcal{B}_c^{\pm}|$).
	(b) Longitudinal resistivity $\rho_{xx}$ and Hall resistivity $\rho_{xy}$ as functions of magnetic field. The data is obtained on a Hall bar of size $L_1\times L_2\times L_3\times W=5\times50\times30\times50$ with a disorder strength $D=0.1$ eV (comparable to the magnetic exchange gap) after averaging over 160 computations. Insets: band structures along the $x$ direction for a 6-SL MBT in its antiferromagnetic (left) and ferromagnetic (right) states.
	  	}
\label{Mag_Hall}
\end{figure}

\textit{In-plane transport properties on a Hall bar.---}
To study the electronic transport, we first discretize the continuum Hamiltonian Eq.~\eqref{eq:Hamiltonian} on a cubic lattice ($a_0=5$nm) invoking the $k\cdot p$ perturbation. Then, under a Hall bar device geometry as illustrated in the inset of Fig.~\ref{Mag_Hall}(a), we calculate the Hall resistivity $\rho_{xy}$ and the longitudinal resistivity $\rho_{xx}$ using the Landauer-B\"uttiker formula~\cite{Supply,datta_1995}. To incorporate fluctuations, we also add a disorder potential $\mathcal{H}_D=V(\bm{r})s_z\otimes\sigma_0$ to the lattice Hamiltonian, where $V(\bm{r})$ is uniformly distributed within $[-D/2,\ D/2]$ with $D$ being the disorder strength. The Fermi level is zero as we do not consider doping or gating.

For a six-SL MBT device reflecting real experimental setup~\cite{Liu2020Robust}, we obtain $\rho_{xx}$ and $\rho_{xy}$ by averaging 160 repeated calculations, which are plotted as functions of magnetic field $\mathcal{B}$ (along $z$) in Fig.~\ref{Mag_Hall}(b). The results show a topological phase transition from a normal insulator (indistinguishable from an axion insulator) with a vanishing Chern number $\mathcal{C}=0$ at low magnetic fields into a quantum anomalous Hall insulator with $\mathcal{C}=\pm 1$ at high magnetic fields. When $\lvert \mathcal{B}\rvert<\mathcal{B}_{c}^+$, the magnetic ground state remains antiferromagnetic with antiparallel spins on adjacent SL, and the system preserves the $\mathcal{PT}$ symmetry. Because the spin flips its sign under $\mathcal{PT}$ operation [$\mathcal{PT}$: $\mathcal{H}(\bm{k},\uparrow)\rightarrow \mathcal{H}(\bm{k},\downarrow)$], the bands must be doubly degenerate with a band gap of $\delta\approx 2\Delta$ at $k_x=0$, as shown in the left inset in Fig.~\ref{Mag_Hall}(b). Consequently, we obtain $\mathcal{C}=0$, hence a vanishing Hall resistivity and a large longitudinal resistivity akin to a normal insulator. While ARPES experiments showed controversial results on the band gap in MBT~\cite{Otrokov2019Prediction,Shikin2020}, transport measurements strongly support the existence of large gaps in both the antiferromagnetic and ferromagnetic states of MBT~\cite{Liu2020Robust,Deng2020QUantum,Ge2020high} by confirming the insulating behavior in longitudinal transport, even though this insulating gap cannot tell axion insulators from normal insulators.

When $\lvert \mathcal{B}\rvert$ exceeds $\mathcal{B}_{c}^+$, however, the magnetic moments undergo a spin-flop transition which breaks the time reversal symmetry for electrons. Correspondingly, the topological Chern number becomes $\mathcal{C}=-\text{sgn}(\mathcal{B})$, leading to a quantized Hall resistivity $\rho_{xy}=h/(\mathcal{C}e^2)$ and a vanishing longitudinal resistivity $\rho_{xx}=0$~\cite{Thouless1982Chern}. The deviations of $\rho_{xy}$ around integer values are ascribed to the finite size effect, which can be suppressed by enlarging the system size. The in-plane resistivities shown in Fig.~\ref{Mag_Hall} agree quantitatively with experimental observations~\cite{Cai2022Electric,Liu2020Robust} widely regarded as evidences for axion insulator. Nevertheless, the topological phase transition taking place here is inadequate to determine an axion insulator because the $\mathcal{C}=0$ phase appearing at small fields by itself is indistinguishable from a normal insulator.

\textit{Surface charge polarization and layer-resolved Chern numbers.---}
A defining feature of axion insulator is the topological TME enabled by the quantized $\theta$-field, which, unlike the Chern number $\mathcal{C}$, can uniquely characterize the axion insulator phase. On the one hand, a magnetic field $\bm{B}$ below the spin-flop threshold will induce a quantized charge polarization $\bm{P}=e^2\theta \bm{B}/(2\pi h)$~\cite{Qi2008Topological}, which is intimately related to the layer-resolved Chern numbers. If the applied $\bm{B}$ field is time dependent, a charge current proportional to $d\bm{P}/dt$ will be generated, enabling a directly detectable signal to be discussed later. On the other hand, the TME also manifests as the magnetization induced by an electric field~\cite{Pournaghavi2021}. However, the TME coefficient quantized by $\theta$ is typically two orders of magnitude smaller than that of ordinary magnetoelectric materials~\cite{nan2008multiferroic}. Therefore, the TME is more amenable to transport measurement as the sensitivity of detecting current is extremely high. Nonetheless, as a consistency check, we also calculated the tiny magnetization induced by an electric field, which indeed turns out to be quantized by the $\theta$ field (see the SM~\cite{Supply}).

\begin{figure}[t]
  \centering
  \includegraphics[width=0.9\linewidth]{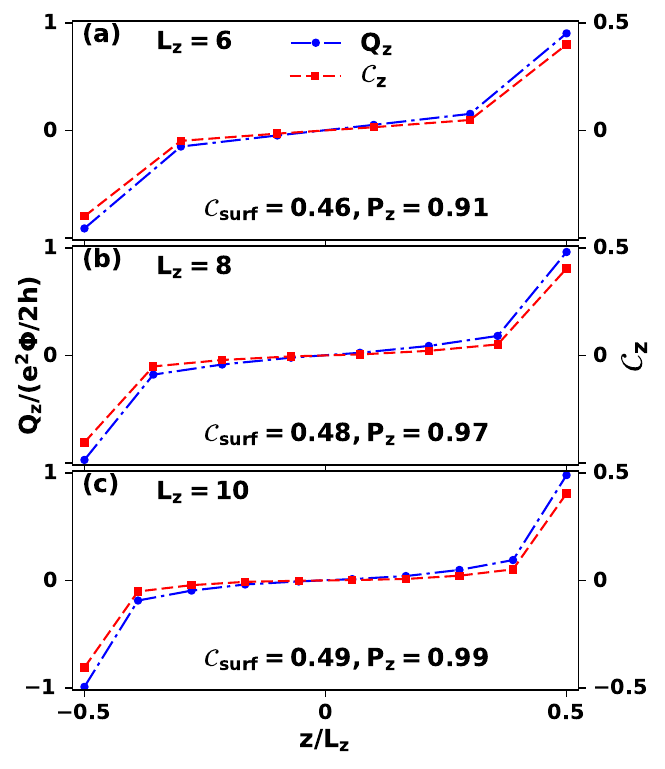}
  \caption{
	Charge distribution among each SL relative to the background average $Q_z-\overline{Q(\bm{r})}$ (blue dots) and layer-resolved Chern numbers $\mathcal{C}_z$ (red squres) for MBT of thickness (a) $L_z=6$, (b) $L_z=8$, and (c) $L_z=10$ on a slab of size $L_x\times L_y\times L_z$ with $L_x=L_y=40$. Here, the magnetic flux per unit cell is $\Phi_0=0.05h/2e$ and the total magnetic flux is $\Phi=L_xL_y\Phi_0$, which corresponds to a magnetic field of $B=2\text{T}$ smaller than the spin-flop threshold $\mathcal{B}_c\approx 3.0\text{T}$ ($a_0=5$\text{nm}).
	  	}
\label{cnum_charge}
\end{figure}

To calculate $\bm{P}$, we consider a slab of thickness $L_z$ and widths $L_x=L_y$ with open boundary conditions and assume that a static magnetic field $\bm{B}=(0,0,B)$ is applied along the $z$-direction, which amounts to a magnetic flux of $\Phi_0=Ba_0^2$ per unit cell. Using the equilibrium Green's function method~\cite{Supply}, we obtain the charge distribution $Q(\bm{r})=-e\langle \hat{n}(\bm{r})\rangle$ where $-e$ is the electron charge and $\hat{n}(\bm{r})$ is the electron density operator. Figure~\ref{cnum_charge} (blue dots) plots the charge distribution among each SL, $Q_z=\sum_{x,y}Q(\bm{r})$, with respect to an averaged background charge $\overline{Q(\bm{r})}=\sum_{x,y,z}Q(\bm{r})/L_z$ which compensates the positive ions in the lattice. Since $Q(\bm{r})$ is an odd function of $z$, as shown in Fig.~\ref{cnum_charge}, there is indeed a finite charge polarization $\bm{P}=\int dV \bm{r}Q(\bm{r})$. As will be shown later, only surface charges contribute to the detectable current, thus only the surface charge polarization $P_z=[Q(L_z/2)-Q(-L_z/2)]/2-\overline{Q(\bm{r})}$ is relevant to our discussion. Ideally, the surface charge polarization $P_z$ should be very close the total polarization $P$, but finite-size effects can bring about deviations. Fortunately, we find that the finite-size effects are well suppressed by increasing the thickness $L_z$. It turns out that $P_z(L_z=6,8,10)=0.91,0.97,0.99$ ($e^2\Phi/2h$) with $\Phi=L_xL_y\Phi_0$ being the total magnetic flux penetrating the slab, rapidly approaching the quantized value determined by the axion field $\theta=\pi$.

We now turn to the layer-resolved Chern numbers $\mathcal{C}_z$ which reflect the relative contribution to the system topology by different SLs. To this end, we adopt periodic boundary conditions in the lateral dimensions under the same slab geometry used above. While the layer-resolved Chern numbers can be straightforwardly obtained by projecting the wavefunctions onto each SL~\cite{Supply} in a clean system, here we resort to the non-commutative approach which is able to incorporate disorders~\cite{Prodan2011Disorder,*Prodan2012Quantum}: $\mathcal{C}_z=-2\pi i\text{Tr}\left\{ \hat{P}\left[ [\hat{x},\ \hat{P}],\ [\hat{y},\ \hat{P}] \right]\hat{P}_z \right\}$, where $\hat{x}$ ($\hat{y}$) is the position operator, $\hat{P}$ is the projector onto the occupied bands, $\hat{P}_z\equiv \lvert\psi_z\rangle\langle\psi_z\rvert$ is the projector onto the $z$-th SL, $[\cdots]$ is the commutator and $\text{Tr}$ denotes the trace. In the presence of $\mathcal{PT}$ symmetry, $\hat{P}_z$ flips sign on opposite surfaces because $\hat{P}_{-z}=|\psi_{-z}\rangle\langle\psi_{-z}|=\mathcal{PT}|\psi_{z}\rangle\langle\psi_{z}|\mathcal{PT}=(\mathcal{PT})^2|\psi_{z}\rangle\langle\psi_{z}|=-|\psi_{z}\rangle\langle\psi_{z}|=-\hat{P}_z$, ensuring that the layer-resolved Chern numbers are odd in $z/L_z$.

Figure~\ref{cnum_charge} shows the layer-resolved Chern numbers $\mathcal{C}_z$ with three different thicknesses $L_z=6,8,10$ (red squares), which agrees remarkably well with the charge distribution $Q_z$. Even in the presence of disorders, we find that $\mathcal{C}_z$ and $Q_z$ are very robust (See SM~\cite{Supply}), suggesting that they are topologically protected properties intrinsic to the axion insulator. Correspondingly, the surface Chern number $\mathcal{C}_{surf}^{L_z}=\sum_{z=-L_z/2}^0\mathcal{C}_z$ is almost half quantized: $\mathcal{C}_{surf}^{6}=0.46$, $\mathcal{C}_{surf}^{8}=0.48$, and $\mathcal{C}_{surf}^{10}=0.49$, indicating a distinct bulk axion field $\theta=\pi$~\cite{Essin2009Magnetoelectric}.

\begin{figure}[t]
  \centering
  \includegraphics[width=0.9\linewidth]{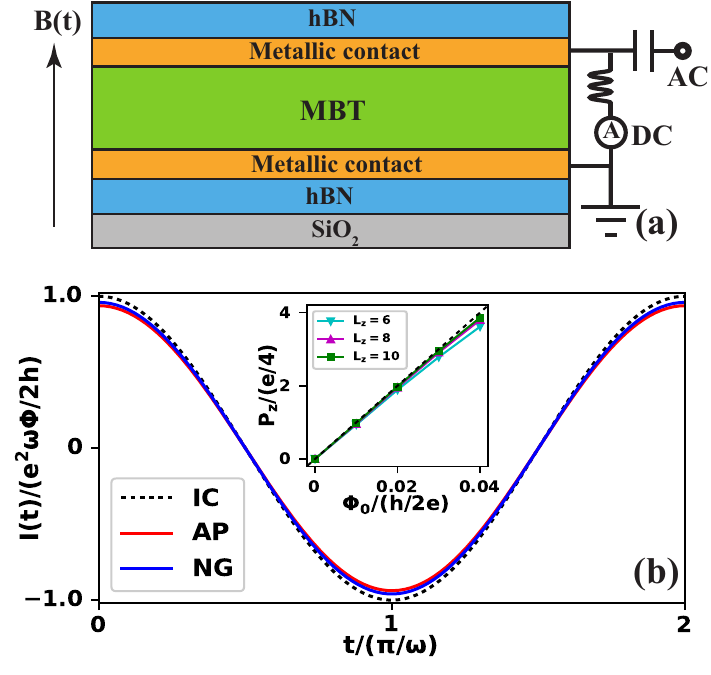}
  \caption{
	(a) Schematics of the proposed AITJ setup to detect the dynamical TME. The metallic leads are connected to a bias tee to separate the harmonic AC output from the DC output which is short circuited. The hexagonal boron nitride (hBN) flakes are added to avoid degradation~\cite{Song2018Giant}, and the whole device is placed on a silicon dioxide substrate.
	(b) Output current $I(t)$ induced by an AC magnetic field $\hat{\bm{z}}B\sin\omega t$ calculated by the time-dependent non-equilibrium Green's function (NG) (solid blue), the adiabatic variation of the surface charge polarization (AP) (solid red), respectively. The ideal case (IC) (dotted black) for an exactly quantized axion field $\theta=\pi$ is plotted as a reference. System size: $L_x=L_y=20$ and $L_z=6$. Magnetic flux per unit cell: $\Phi_0=0.05h/2e$. Driving frequency: $\hbar\omega=0.001\text{eV}$. Inset: surface charge polarization as a function of the magnetic flux for three different thicknesses $L_z=6,8,10$.
	  	}
\label{MTJcurrent}
\end{figure}

\textit{Charge current in MBT tunnel junction.---}
To detect the quantized TME in MBT using transport experiment, we need to consider a time-dependent magnetic field such that the induced surface charge polarization becomes dynamical and produces a charge current in the $z$ direction. This approach has been utilized to characterize multiferroic materials exhibiting non-quantized magnetoelectric effects~\cite{nan2008multiferroic}.
To this end, we conceive an axion insulator tunnel junction (AITJ) consisting of an even-SL MBT sandwiched between two metallic contacts~\footnote{In artificial heterostructures involving 2D materials, graphene contacts are preferred (see, for example, Ref.~\cite{Song2018Giant,Jiang2018}) because it can be incorporated without the need for lattice matching.} as illustrated in Fig.~\ref{MTJcurrent}(a). In the adiabatic limit, {the charge polarization follows the magnetic field at any instant of time}, which can be detected directly as an AC output signal from the AITJ. Since the metallic contacts are connected only to the top and bottom layers, only the surface polarization $P_z$ is relevant to the transport measurement.

Using the lattice Hamiltonian, we resort to the time-dependent non-equilibrium Green's function to compute the output current in the AITJ~\cite{Supply}. A harmonic magnetic field $\bm{B}(t)=\hat{\bm{z}}B\sin{\omega t}$ applied to the AITJ converts to a phase $\Phi_0(t)=\Phi_0\sin{\omega t}$ for electrons, where $\Phi_0=Ba_0^2$ is the magnetic flux per unit cell. As a result, the effective Hamiltonian acquires a time-dependent perturbation that drives the electron motion, forming a charge current. Since the system is now periodic in time, the induced charge current can be expanded into a Fourier series as~\cite{Supply,Li2018Doubled}
\begin{align}
I(t)=\sum_{n=-\infty}^{\infty}I_ne^{in\omega t},
\end{align}
where $I_n$ is $n$-th harmonic component satisfing $I_n=-I_{-n}^*$, ensuring a real current. The total current $I(t)$ includes a DC component $I_0$ and a series of AC components $I_{n>0}$. Truncating the Green's function at order $n=4$ suffices to yield a converging result~\cite{Supply}. Figure~\ref{MTJcurrent}(b) (solid blue curve) plots the numerical result of $I(t)$ for one period of oscillation, where the first order term $I_{n=\pm 1}$ indeed dominates all other components. The plot is offset by $I_0$ because this DC component is short circuited via the bias tee illustrated in Figure~\ref{MTJcurrent}(a).

As an independent confirmation, we use the same harmonic field $\bm{B}(t)=\hat{\bm{z}}B\sin{\omega t}$ in the surface charge polarization $P_z$ and calculate the resulting charge current $I_z(t)=dP_z(t)/dt$~\cite{Li2010Dynamical}, assuming an adiabatic condition that $\bm{P}$ undergoes a quasi-static variation without inter-band transitions induced by the oscillating $\bm{B}(t)$~\cite{adiabatic}. Figure~\ref{MTJcurrent}(b) (solid red curve) plots $I_z(t)$ within one period of oscillation for a same MBT slab, which agrees remarkably well with the harmonic signal $I(t)$ obtained by the non-equilibrium Green's function method. To benchmark the accuracy of our numerical results, we also plot the ideal case for an infinite system, where $I(t)=dP(t)/dt=\theta e^2/(2\pi h)\omega\Phi\cos\omega t$ with a strictly quantized axion field $\theta=\pi$ (dotted black curve in Fig.~\ref{MTJcurrent}(b)). We see that our numerical results obtained both from the non-equilibrium Green's functions and from $P_z(t)$ only slightly deviate from the ideal case, which demonstrates the validity and reliability of our proposal. We mention in passing that if the Fermi level is tuned into the conducting band (\textit{e.g.}, by gating the device~\cite{Jiang2018}), the MBT will become metallic and the induced AC current will vanish.

For an MBT of size $L_x=L_y=10\rm{\mu m}$, a harmonic magnetic field of strength $B\sim100$Gs and frequency $\omega/2\pi=1$GHz induces an output AC current $I\sim121.54$nA, which is a conservative estimation. Since $I$ scales as $\omega B L_xL_y$, the output current can be amplified by increasing the driving frequency $\omega$, the magnetic field $B$, or the system size in the lateral dimensions. In the ideal case, the induced surface charge polarization $P_z$ should scale linearly with the magnetic flux per unit cell $\Phi_0$. To evaluate potential deviations due to finite-size effects, we plot $P_z$ as a function of $\Phi_0$ for different thicknesses against the ideal scaling in the inset of Fig.~\ref{MTJcurrent}(b), where the finite-size effects turn out to be negligible, further confirming the validity of our calculations.

In summary, we have theoretically proposed an experimental setup to unambiguously identify antiferromagnetic MBT as an axion insulator by detecting the AC current induced by a harmonic magnetic field under the adiabatic condition. Comparing to the vanishing Hall resistance measured in previous experiments, which is inadequate to confirm the axion insulator phase, our proposed scheme provides a smoking-gun signal to identify MBT as an axion insulator.

\textit{Acknowledgments.---} This work is supported by the Air Force Office of Scientific Research under grant FA9550-19-1-0307. We acknowledge helpful discussions with Chong Wang and U. K. R\"o\ss{}ler.

\bibliography{ref}

\begin{thebibliography}{49}
\bibitem [{adi()}]{adiabatic}%
  \BibitemOpen
  \href@noop {} {}\bibinfo {note} {So long as the driving frequency
  $\hbar\omega$ is far less than the band gap $\Delta$ of the MBT, we can treat
  the magnetic field as varying adiabatically. The driving frequency of the
  magnetic field $\hbar\omega$ is on the order of $\mu$eV, which is indeed far
  less than the energy gap $\Delta\sim 100$ meV.}\BibitemShut {Stop}%
\end{thebibliography}

\end{document}


\title{Supplemental material for ``Identifying Axion Insulator by Quantized Magnetoelectric Effect in antiferromagnetic ${\mathrm{MnBi}}_{2}{\mathrm{Te}}_{4}$ Tunnel Junction"}

\author{Yu-Hang Li}
\email[]{yuhang.li@ucr.edu}
\affiliation{Department of Electrical and Computer Engineering, University of California, Riverside, California 92521, USA}
\author{Ran Cheng}
\email[]{rancheng@ucr.edu}
\affiliation{Department of Electrical and Computer Engineering, University of California, Riverside, California 92521, USA}
\affiliation{Department of Physics and Astronomy, University of California, Riverside, California 92521, USA}

\date{\today}
\maketitle

\section{I. Lattice Hamiltonian}
We use the $k\cdot p$ theory to discretize the effective Hamiltonian for MnBi\textsubscript{2}Te\textsubscript{4} (MBT) on a cubic lattice $\psi_{\bm{i}}=\begin{pmatrix}c_{\bm{i}\uparrow}^s,&&c_{\bm{i}\uparrow}^p,&&c_{\bm{i}\downarrow}^s,&&c_{\bm{i}\downarrow}^p \end{pmatrix}^T$ with $c_{\bm{i}\sigma}^{s(p)}$ the annihilating operator for an electron with orbital $s$ ($p$) spin $\sigma$ on site $\bm{i}=\begin{pmatrix}
x,&y,&z
\end{pmatrix}$. Substituting the momentum $k_{\alpha=x,y,z}=-i\partial_{\alpha}\rightarrow-i(\psi_{\bm{i}+\alpha}-\psi_{\bm{i}-\alpha})/(2a_0)$ and $k_{\alpha}^2=-\partial_{\alpha}^2\rightarrow-(\psi_{\bm{i}+\alpha}+\psi_{\bm{i}-\alpha}-2\psi_{\bm{i}})/a_0^2$ with $a_0$ the lattice constant, we can write the lattice Hamiltonian as 
\begin{align}
  \mathcal{H}=\sum_{\bm{i}}\psi_{\bm{i}}^{\dagger}(\bm{T}_{\bm{i}}+\bm{T}_{zM}+\bm{T}_D)\psi_{\bm{i}}+(\psi_{\bm{i}}^{\dagger}\bm{T}_{x}\psi_{\bm{i}+x}+\psi_{\bm{i}}^{\dagger}\bm{T}_{y}\psi_{\bm{i}+y}+\psi_{\bm{i}}^{\dagger}\bm{T}_{z}\psi_{\bm{i}+z}+\text{H.c.}),
  \label{Lattic_Hamiltonian}
\end{align}
where
\begin{align}
&\bm{T}_{\bm{i}}=(M_0+2B_1+4B_2)s_{0}\otimes\sigma_z,\notag \\ \notag
&\bm{T}_{zM}=\Delta\bm{m}(z)\cdot\bm{s}\otimes\sigma_0,\\  \notag
&\bm{T}_{D}=D(\bm{i})\cdot s_z\otimes\sigma_0,\\  \notag
&\bm{T}_{x}=-B_2s_0\otimes\sigma_z-iA_2/(2a_0)s_x\otimes\sigma_x,\\  \notag
&\bm{T}_{y}=-B_2s_0\otimes\sigma_z-iA_2/(2a_0)s_y\otimes\sigma_x,\\  \notag
&\bm{T}_{z}=-B_1s_0\otimes\sigma_z-iA_1/(2a_0)s_z\otimes\sigma_x.  \notag
\end{align}
Here, $\bm{T}_{\bm{i}}$ is the onsite term. $\bm{T}_{\alpha=x,y,z}$ is the hoping term along the $\alpha$ direction. $\bm{T}_{zM}$ is the exchange coupling between topological electrons and magnetic moments. $\Delta$ is the exchange gap. $\bm{m}_{z}=\begin{pmatrix}\sin{\theta_{z}},&0,&\cos{\theta_z}\end{pmatrix}$ with $\theta_z$ the polar angle of the magnetic moment in $z$-th septuple layer, which is assumed to be localized inside the $xz$ plane for simplicity. $\bm{T}_{D}$ represents magnetic disorders, where $D(\bm{i})$ is a random number within $[-D/2,\ D/2]$ with $D$ the disorder strength. Other parameters are defined in the main text.
 
In the presence of a static magnetic field $\bm{B}=\begin{bmatrix}0,&0,&B\end{bmatrix}$, the lattice Hamiltonian acquires a phase factor $\phi_{\bm{ij}}=2\pi\int_{\bm{i}}^{\bm{j}}\bm{A}\cdot d\bm{r}/\phi_0$ where $\phi_0=h/(2e)$ is the magnetic flux quanta and $\bm{A}$ is the vector potential ($\nabla{\times\bm{A}}=\bm{B}$) Here, we adopt the Landau gauge $\bm{A}=\begin{bmatrix}yB,&0,&0\end{bmatrix}$. Using the Peierls substitution, we can recasts the lattice Hamiltonian into
 \begin{align}
   \mathcal{H}=\sum_{\bm{i}}\psi_{\bm{i}}^{\dagger}(\bm{T}_{\bm{i}}+\bm{T}_{zM})\psi_{\bm{i}}+[\psi_{\bm{i}}^{\dagger}\bm{T}_{x}f_x(y)\psi_{\bm{i}+x}+\psi_{\bm{i}}^{\dagger}\bm{T}_{y}\psi_{\bm{i}+y}+\psi_{\bm{i}}^{\dagger}\bm{T}_{z}\psi_{\bm{i}+z}+\text{H.c.})],
  \label{Lattic_Hamiltonian_phase}
\end{align}
where $f_x(y)=e^{i\phi_{\bm{ii}+x}}=e^{iyBa_0^2/\phi_0}$.
 
Under a time dependent magnetic field $\bm{B}=\begin{bmatrix}0,&0,&B(t)\end{bmatrix}$, the vector potential $\bm{A}(t)$ becomes time dependent, so is the $f_x(y,t)$ term. In particular, if the magnetic field is harmonic, \textit{i.e.}, $B(t)=B_0\sin{\omega t}$, the lattice Hamiltonian is also periodic $\mathcal{H}(t)=\mathcal{H}(t+T)$ with $T=2\pi/\omega$, which can be written as $\mathcal{H}(t)=H_0+H(t)$ on a slab of size $L_x\times L_y \times L_z$, where $H_0$ is the time independent component while $H(t)=\sum_{\bm{i}}(\psi_{\bm{i}}^{\dagger}T_xe^{ipB_0a_0^2\sin{\omega t}}\psi_{\bm{i}+x}+\text{h.c.})$. Using the expansion $\exp{[iz\sin{(\alpha)}]}=\sum_{k=-\infty}^{\infty}\mathcal{J}_{k}(z)\exp{(ik\alpha)}$~\cite{Gradshteyn}, we can finally rewrite the lattice Hamiltonian as 
\begin{align}
    \mathcal{H}(t)=H_0+\sum_{k=-\infty}^{\infty}\sum_{\bm{i}}[\psi_{\bm{i}}^{\dagger}T_x\mathcal{J}_{k}(y\Phi_0)\exp{(ik\omega t)}\psi_{\bm{i}+x}+\psi_{\bm{i}+x}^{\dagger}T_x^\dagger\mathcal{J}_{k}(-y\Phi_0)\exp{(ik\omega t)}\psi_{\bm{i}}], \label{eq:Ht}
\end{align}
where $\mathcal{J}_{k}(\pm y\Phi_0)$ is Bessel function of the first kind at order $k$, and $\Phi_0=B_0a_0^2$ is the magnetic flux per unit cell. When setting $\Phi_0=0$ (removing the magnetic field), Equation~\eqref{eq:Ht} reduces to Eq.~\eqref{Lattic_Hamiltonian_phase} because $\mathcal{J}_{k}(0)=\delta_{k}$ and $\sum_k\mathcal{J}_k(x\ne0)=1$.

\section{II. Magnetic configuration}
\begin{figure}[t]
  \centering
  \includegraphics[width=0.6\linewidth]{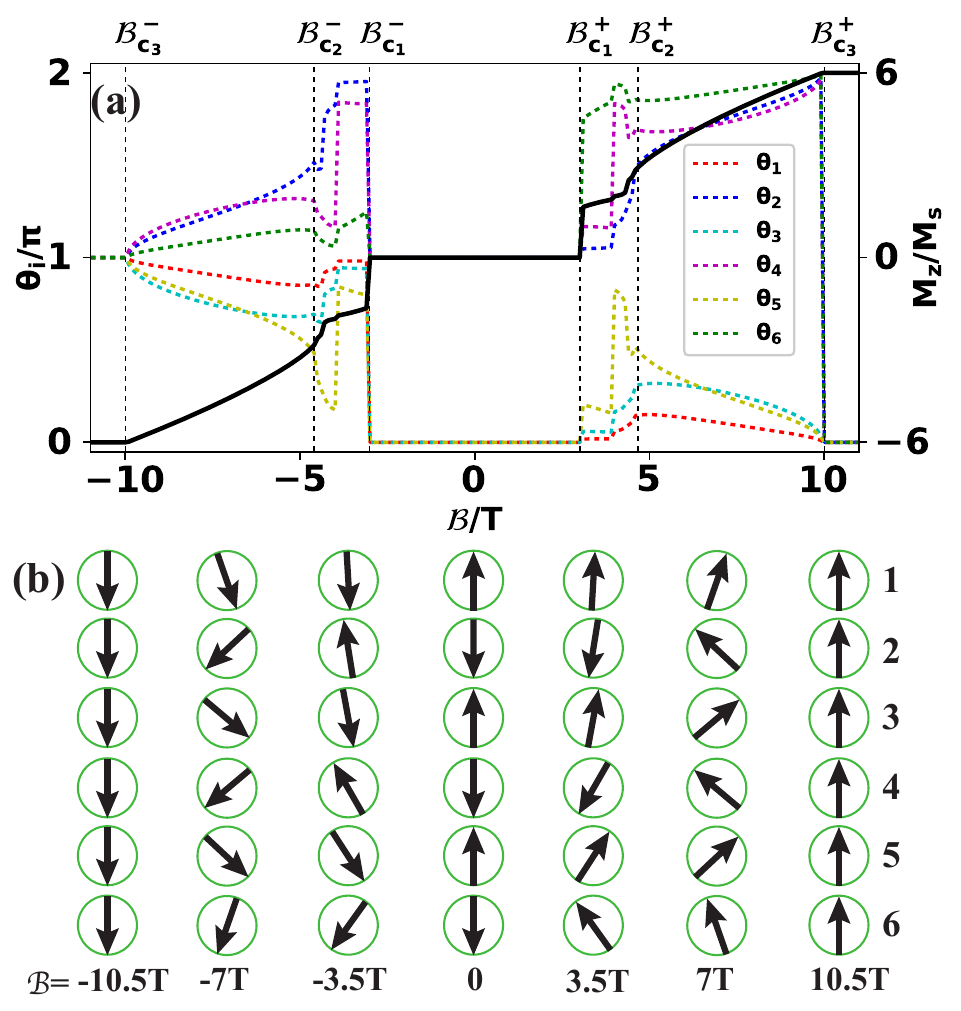}
  \caption{
  	Illustration of magnetic configuration for a six-septuple layer MBT.
  	(a) Evolution of the equilibrium magnetization orientation $\theta_i$ and total magnetization $M_z$ as functions of the applied magnetic field $\mathcal{B}$. The critical fields marking configurational transitions are labeled on the top.
  	(b) Schematic magnetic configurations at different magnetic fields, which are representative of seven different magnetic states. The parameters adopted are shown in Table I in the main text.
  	}
\label{Mag}
\end{figure}
The magnetic properties of a N-septuple layer MBT can be captured by the free energy shown in Eq. 2 in the main text. Minimizing the free energy gives the equilibrium magnetic configurations. Figure~\ref{Mag} (a) shows the magnetic configuration and the total magnetization as functions of external magnetic field $\mathcal{B}$ for a six-septuple layers= MBT. In general, the system has seven distinct magnetic states as the magnetic field is tuned continuously from $-11$ Tesla to $11$ Tesla. When $\lvert \mathcal{B}\rvert<\mathcal{B}_{c_1}^+$, the Zeeman energy is not large enough to overcome the anisotropy. Consequently, the system stays in its antiferromagnetic ground state with a vanishing magnetization. Namely, the spins in adjacent layers are antiparallel to each other,  preserving the combined parity and time reversal ($\mathcal{PT}$) symmetry. For $\mathcal{B}_{c_1}^+<\lvert \mathcal{B}\rvert<\mathcal{B}_{c_2}^+$, the system turns into a canted state, in which the canting angles among different layers are somehow random. Meanwhile, the magnetization exhibits a sudden change at the critical point $\mathcal{B}_{c_1}^{\pm}$ where the Zeeman energy exceeds the magnetic anisotropy. For $\mathcal{B}_{c_2}^+<\lvert \mathcal{B}\rvert<\mathcal{B}_{c_3}^+$, the system enters a spin flop state with the canting angle of layer $z$ and layer $-z$ satisfying $\theta_{z}+\theta_{-z}=2\pi$. The system finally becomes ferromagnetic with all spins collinear with the magnetic field when $\lvert \mathcal{B}\rvert>\mathcal{B}_{c_3}^+$. Figure.~\ref{Mag}(b) schematically illustrates the seven different spin configurations taking place at different magnetic fields. 

\section{III. Landauer B\"uttiker formulism}
With the discretized Hamiltonian, we can calculate the Hall resistivity and longitudinal resistivity in a Hall-bar device geometry. The Landauer B\"uttiker formula defines the transmission coefficient from lead $n$ to lead $m$ as
\begin{align}
T_{mn}=\text{Tr}(\Gamma_mG^r\Gamma_nG^a),
\end{align}
where $G^{r,a}=(-\mathcal{H}-\sum_{m=1}^6\Sigma_m^{r,a})^{-1}$ are the retarded and advance Green's functions. $\Gamma_m=i(\Sigma_m^r-\Sigma_m^a)$, with $\Sigma_m^{r,a}$ the self energy, is the line width function. The potential $V_m$ of lead $m$ can then be determined by solving
\begin{align}
\begin{bmatrix}T_{11}&T_{12}&T_{13}&T_{14}&T_{15}&T_{16}\\T_{21}&T_{22}&T_{23}&T_{24}&T_{25}&T_{26}\\T_{31}&T_{32}&T_{33}&T_{34}&T_{35}&T_{36}\\T_{41}&T_{42}&T_{43}&T_{44}&T_{45}&T_{46}\\T_{51}&T_{52}&T_{53}&T_{54}&T_{55}&T_{56}\\T_{61}&T_{62}&T_{63}&T_{64}&T_{65}&T_{66}\end{bmatrix}\times\begin{bmatrix}V_1\\V_2\\V_3\\V_4\\V_5\\V_6\end{bmatrix}=\begin{bmatrix}I_1\\I_2\\I_3\\I_4\\I_5\\I_6\end{bmatrix},
\end{align}
where $T_{ii}=-\sum_{j\ne i}T_{ij}$, and the current vector is set to be $\hat{I}=\begin{bmatrix}I,&0,&0,&-I,&0,&0\end{bmatrix}$. The Hall resistivity $\rho_{xy}=(V_2-V_6)/I$ and the longitudinal resistivity  $\rho_{xx}=(V_2-V_3)/I$ in the main text are obtained by averaging  160 repeated numerical calculations.

\section{IV. Layer-resolved Chern numbers using the TKNN formula}

\begin{figure}[b]
  \centering
  \includegraphics[width=0.5\linewidth]{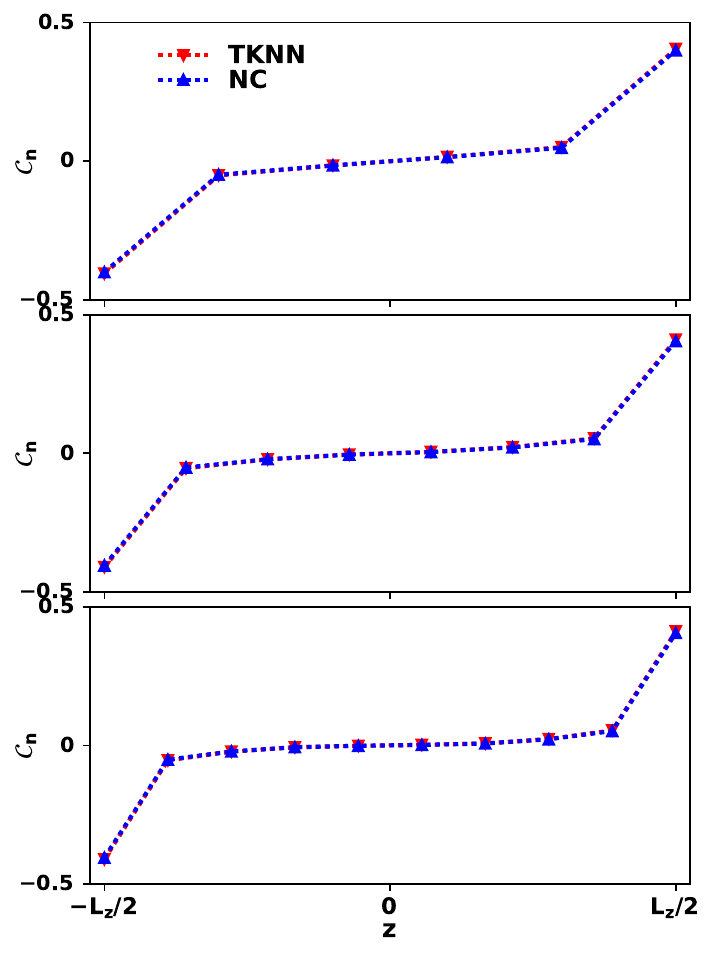}
  \caption{
  Layer-resolved Chern numbers calculated using the TKNN formula (red triangle) and the non-commutative formula (blue triangle) for three different thicknesses (a )$L_z=6$, (b) $L=8$ and (c) $L=10$. The results coincide with each other. The lateral dimensions of the slab are chosen to be $L_x=L_y=40$ when computing with the non-commutative formula.
  	}
\label{TKNNcn}
\end{figure}

The layer-resolved Chern numbers can be alternatively obtained using the TKNN formula~\cite{Thouless}
\begin{align}
    \mathcal{C}_z=\frac{1}{2\pi}\int{d\bm{k}^2}\sum_{\epsilon_m<\epsilon_F<\epsilon_n}\frac{\langle m |\hat{P}_z\frac{\partial{\mathcal{H}}}{\partial{k_x}}|n\rangle\langle n|\frac{\partial{\mathcal{H}}}{\partial{k_y}}|m\rangle-\text{H.c.}}{[\epsilon_m(\bm{k})-\epsilon_n(\bm{k})]^2},
    \label{TKNNcnum}
\end{align}
where $\epsilon_F=0$ as the Fermi energy is inside the surface gap, $\epsilon_{m(n)}(\bm{k})$ and $|m\rangle$ ($|n\rangle$) are the eigenenergy and eigenstate of $\mathcal{H}$, $\hat{P}_z\equiv|\psi_z\rangle\langle\psi_z|$ is the projector onto $z$-th layer. Figure.~\ref{TKNNcn} shows the layer-resolved Chern numbers calculated using Eq.~\ref{TKNNcnum} and the non-commutative formula for three different thicknesses $L_z=6,8,10$ of width $L_x=L_y=40$. The two approaches agree remarkably well with each other.

\section{V. Charge density}
\begin{figure}[ht]
  \centering
  \includegraphics[width=0.55\linewidth]{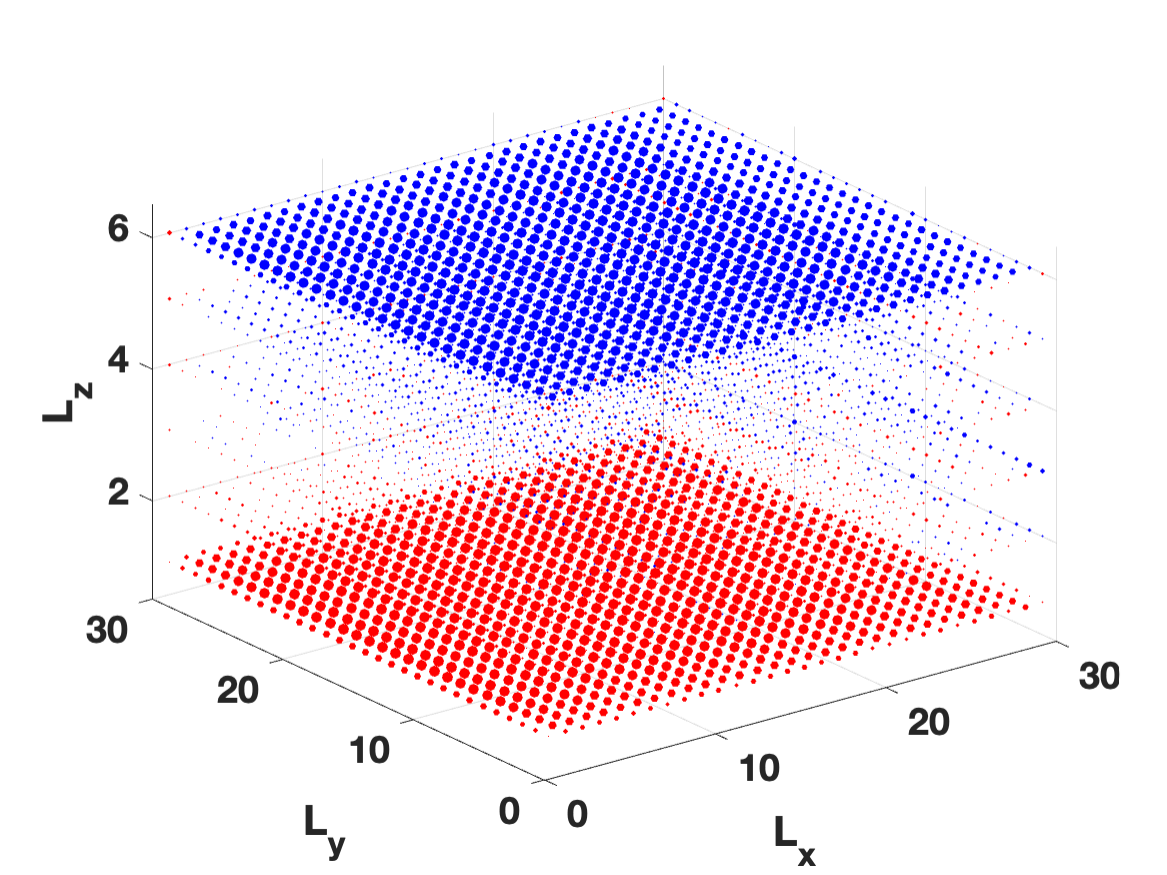}
  \caption{
  	Polarized charge distribution on a MBT slab of size $L_x\times L_y\times L_z=30\times 30\times 6$ under a magnetic flux $\Phi_0=0.01h/(2e)$. Here, the blue (red) dots represent positive (negative) charges, the size of which reflects the local charge density.
  	}
\label{fig:chargepolarization}
\end{figure}

In the presence of a static magnetic field $\bm{B}=(0,0,B)$, the charge density can be expressed as
\begin{align}
Q(\bm{r})=-e\langle\hat{n}(\bm{r})\rangle=-e\langle\psi^\dagger(\bm{r},t)\psi(\bm{r},t)\rangle=-ie\text{Tr}[G^<(\bm{r},t;\bm{r}^\prime,t)],
\end{align}
where $-e$ is the electron charge, $\hat{n}(r)=\psi^\dagger(\bm{r},t)\psi(\bm{r},t)$ is the number operator, $G^<(\bm{r},t;\bm{r}^\prime,t)$ is the distribution Green's function and $\text{Tr}$ represents the trace of a matrix. After a Fourier transform, we have
\begin{align}
Q(\bm{r})=-ie\int{d\epsilon}\text{Tr}[G^<(\bm{r},\bm{r}^\prime;\epsilon)],
\label{eq:charge}
\end{align}
where the distribution Green's function $G^<(\bm{r},\bm{r}^\prime;\epsilon)=f(\epsilon)[G^a(\bm{r},\bm{r}^\prime;\epsilon)-G^r(\bm{r},\bm{r}^\prime;\epsilon]$. Here, the Fermi distribution function $f(\epsilon)=1$ when $\epsilon\le 0$ while $f(\epsilon)=0$ when $\epsilon>0$ (at $T=0$K), and the Green's function $G^{r(a)}(\bm{r},\bm{r}^\prime;\epsilon)=1/[\epsilon\pm i\gamma-\mathcal{H}]$ where $\gamma$ is the imaginary line width function. Figure~\eqref{fig:chargepolarization} plots the $Q(\bm{r})$ calculated from Eq.~\eqref{eq:charge} in the thickness direction, where we see that induced charges essentially locates on the top and bottom layers.

\section{VI. Disorder effect}
\begin{figure}[ht]
  \centering
  \includegraphics[width=0.55\linewidth]{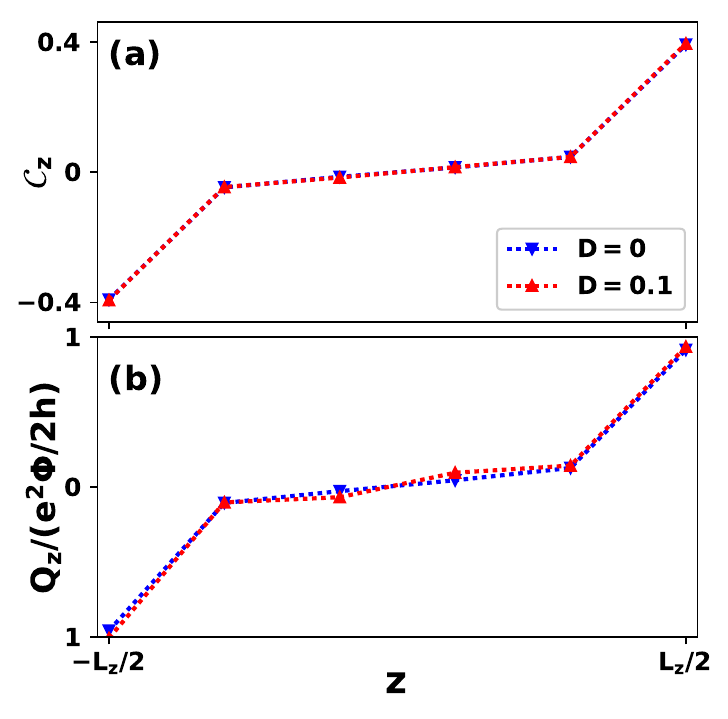}
  \caption{
  	(a) Layer-resolved Chern number on a MBT slab under different disorder strength $D=0$ and $0.1\text{eV}$. 
	(b) Layer-resolved charges induced by a static magnetic filed of strength $\Phi_0=0.01\phi_0$ along the $z$-direction.
	Here, the system size is $L_x\times L_y\times L_z=30\times 30\times 6$.
  	}
\label{disorder}
\end{figure}

When fabricating the experimental setup proposed in the main text, strong magnetic disorders may be present. It is thus imperative to study to what extend the magnetic disorders can affect the axion insulator state of MBT. To this end, we calculate the layer-resolved Chern number $\mathcal{C}_n$ and layer-resolved polarized charge $P_z$ under magnetic disorders $D=0.1\text{eV}$, which is strong enough to collapse the surface band gap. As a comparison, we also consider the case without disorders, and show the results together in Fig.~\ref{disorder}. Overall, both the layer-resolved Chern numbers and the polarization remain almost unchanged in the presence of strong disorders. Specifically, the surface Chern numbers are $\mathcal{C}_{surf}^6(D=0)=0.4524$ and $\mathcal{C}_{surf}^6(D=0.1)=0.4547$, which are very close to each other and almost invisible in Fig.~\ref{disorder}(a). The surface polarization is $P_z(D=0)=0.9362  e^2\Phi/2h$ and  $P_z(D=0.1)=0.9713 e^2\Phi/2h$, where the disorder-induced change is only $3\%-4\%$. These results clearly demonstrate that the axion insulator phase is robust against magnetic disorders, which is consistent with previous studies~\cite{Li,Song}.

\section{VII. Electric field induced magnetization}

The magnetoelectric response manifests in two complementary aspects. While in the main text we focus on the charge polarization induced by a magnetic field, here, for consistency, we check the magnetization induced by an electric field. For $\bm{E}=(0,\ 0,\ E)$, the Fermi level experiences a layer-dependent perturbation. The Fermi energy of layer $z$ is thus $\mu_F(z)=(z-1)/L_z\delta\mu_F$ where $\delta\mu_F=-eEL_z$ is the total Fermi energy drop. Since this electric field does not break the in-plane transnational symmetry, we can thus resort to the following equation to calculate the magnetization in a unit cell~\cite{Thonhauser2005,Pournaghavi2021}
\begin{align}
M=\frac{e}{2h}\sum_{n}\text{Im}\int_{1^{st} BZ}{\frac{dk^2}{2\pi}}\langle\partial_{\bm{k}}u_{n\bm{k}}|\times(E_{n\bm{k}}+\hat{H}_{\bm{k}})|\partial_{\bm{k}}u_{n\bm{k}}\rangle,
\label{mag_eq}
\end{align} 
where $\hat{H}_{\bm{k}}$ is the lattice Hamiltonian in the momentum space, $E_{n\bm{k}}$ is the energy of the $n$-th band, and $u_{n\bm{k}}$ is the corresponding Bloch wave function. The integral is performed within the first Brillouin zone and the summation is taken over the occupied bands. In Eq.~\ref{mag_eq}, the magnetization is scaled into the unit of $e/2h\cdot\text{meV}$. 

\begin{figure}[ht]
  \centering
  \includegraphics[width=0.8\linewidth]{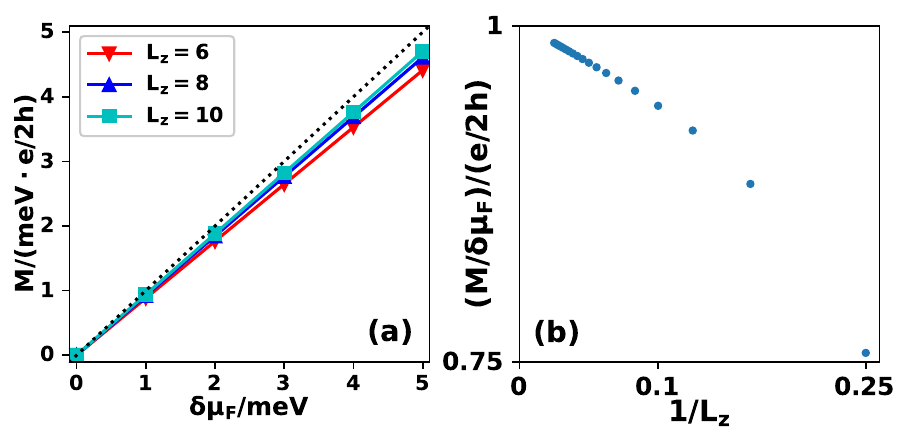}
  \caption{
	(a) Magnetization per unit cell as a function of the total Fermi energy drop (proportional to the electric field) for layer thicknesses $L_z=6,8,10$. The solid lines are guides to the eyes and the black dotted line is the ideal case with $\theta=\pi$ (or, $M=e^2/(2h)EL_za_0=e/(2h)\delta\mu_F$).
	(b) Magnetization as a function of the inverse of the layer thickness from $L_z=4$ to $L_z=40$ with a total electric-field-induced Fermi energy drop $\delta\mu_F=1$ meV.
	All parameters are exactly the same with those taken in the main text.
	  	}
\label{mag}
\end{figure}

Figure.~\ref{mag} shows the electric field induced magnetization for ${\mathrm{MnBi}}_{2}{\mathrm{Te}}_{4}$ under exactly the same parameters used in the main text. In general, the results here independently confirm the quantized topological magnetoelectric effect in axion insulators, which, within the computational accuracy, equals to $e/2h$ when the thickness approaches infinite.

\section{VIII. Non-equilibrium Green's function method for the time dependent current in the AITJ junction}
The full Hamiltonian for the axion insulator tunnel junction shown in Fig. 3 in the main text is
\begin{align}
\mathcal{H}_{full}(t)=\sum_{\alpha=u,p}\mathcal{H}_{\alpha}+\mathcal{H}_{C}(t)+\mathcal{H}_{T},
  \label{Full_Hamiltonian}
\end{align}
where $\mathcal{H}_{\alpha=u(d)}$ represents the Hamiltonian of the upper (lower) metallic contact, $\mathcal{H}_{C}(t)=H_0+H(t)$ is the time dependent Hamiltonian of MnBi\textsubscript{2}Te\textsubscript{4} under a harmonic magnetic field and $\mathcal{H}_{T}$ describes the coupling between metallic contacts and the central region. Conveniently, the metallic contacts can be easily realized by tuning the Fermi surface of the axion Hamiltonian into the conducting band.

The current from the upper contact into the central region can be denoted as
\begin{align}
I_u(t)&=-e\langle\frac{d\hat{N}_u(t)}{dt}\rangle=\frac{ie}{\hbar}\langle[\hat{N}_u(t),\ \mathcal{H}_{full}(t)]\rangle\\
&=\frac{e}{\hbar}\text{Tr}[G_{Cu}^{<}(t,t)\hat{T}_{uC}-\hat{T}_{Cu}G_{uC}^{<}(t,t)],
\label{current_def}
\end{align}
where $\hat{N}_u(t)$ is the particle number in lead $u$. $\hat{T}_{uC}$($=\hat{T}_{Cu}^\dagger$) is the coupling matrix between contact $u$ and the central region. $G_{Cu}^{<}(t,t)$ and $G_{uC}^{<}(t,t)$ are lesser Green's functions. We have used the Heisenberg equation of motion in deriving Eq.~\ref{current_def}. Since the Hamiltonian is time dependent and periodic, we can thus perform the double Fourier transform $C^<_{uC(Cu)}(t,t^\prime)=\frac{1}{2\pi}\int{d\epsilon}{d\epsilon^\prime}e^{-i\epsilon t/\hbar}e^{-i\epsilon^\prime t^\prime/\hbar}C^<_{uC(Cu)}(\epsilon,\epsilon^\prime)$, which recasts the current as
\begin{align}
I_u(t)=\frac{e}{h}\int{d\epsilon}{d\epsilon^\prime}e^{-i(\epsilon-\epsilon^\prime)t/\hbar}\text{Tr}[G_{Cu}^<(\epsilon,\epsilon^\prime)\hat{T}_{uC}-\hat{T}_{Cu}G_{uC}^<(\epsilon,\epsilon^\prime)].
\end{align}
As will be shown later, the Green's function $G_{Cu(uC)}^<(\epsilon,\epsilon^\prime)$ is nonzero only when $\epsilon-\epsilon^\prime=n\hbar\omega$ where $\omega$ is the magnetic frequency and $n$ is an integer, \textit{i.e.}, $G_{Cu(uC)}^<(\epsilon,\epsilon^\prime)=\sum_n G_{Cu(uC)}^<(\epsilon,\epsilon+n\hbar\omega)\delta(\epsilon^\prime-\epsilon-n\hbar\omega)$. Consequently, the current can be further written as
\begin{align}
I_u(t)=\frac{e}{h}\sum_n\int{d\epsilon}e^{in\omega t}\text{Tr}[G_{Cu}^<(\epsilon,\epsilon+n\hbar\omega)\hat{T}_{uC}-\hat{T}_{Cu}G_{uC}^<(\epsilon,\epsilon+n\hbar\omega)].
\end{align}
Equivalently,
\begin{align}
I_u(t)=\frac{e}{h}\sum_n\int{d\epsilon}e^{in\omega t}\text{Tr}[G_{Cu}^<(\epsilon,\epsilon+n\hbar\omega)\hat{T}_{uC}-\hat{T}_{Cu}G_{uC}^<(\epsilon-n\hbar\omega,\epsilon)].
\end{align}
For convenience, we define $G^<_{uC(Cu),mn}(\epsilon)=G^<_{uC(Cu)}(\epsilon+m\hbar\omega,\epsilon+n\hbar\omega)$. In this form, we finally obtain
\begin{align}
I_u(t)=\frac{e}{h}\sum_n\int{d\epsilon}e^{in\omega t}\text{Tr}[G_{Cu,0n}^<(\epsilon)\hat{T}_{uC}-\hat{T}_{Cu}G_{uC,-n0}^<(\epsilon)].
\label{current_final}
\end{align}
Equation.~\ref{current_final} is the central formula to calculate the time dependent current induced by a harmonic magnetic field in an axion insulator, from which we notice that the current $I(t)=\sum_nI_ne^{in\omega t}$ where $I_n=i_n+i_{-n}^*$ with $i_n=\frac{e}{h}\int{d\epsilon}\text{Tr}[G_{Cu,0n}^<(\epsilon)\hat{T}_{uC}]$ and $i_{-n}^*=-\frac{e}{h}\int{d\epsilon}\text{Tr}[\hat{T}_{Cu}G_{uC,-n0}^<(\epsilon)]$.

The question is now simplified to calculate the non-equilibrium Green's functions $G_{mn}^{<(r,a)}(\epsilon)$. Let us first calculate the retarted Green's function. Starting from the full Hamiltonian shown in Eq.~\ref{Full_Hamiltonian}, the retarted Green's function $G^r(t,t^{\prime})$ is defined as
\begin{align}
G^r(t,t^{\prime})=-i\theta(t-t^\prime)\sum_{\bm{ij}}\langle\{\Psi_{\bm{i}}(t), \Psi_{\bm{j}}^{\dagger}(t^\prime)\}\rangle,
\end{align}
where $\theta(x)$ is the step function, $\Psi_{\bm{i}}(t)$ [$\Psi_{\bm{j}}^{\dagger}(t^\prime)$] is the annihilation (creation) operator for electron in the system, $\{\cdots\}$ denotes the anticommutator and $\langle\cdots\rangle$ is the state average on the basis of $\mathcal{H}_{full}(t)$.  Perform the equation of motion, we obtain
\begin{align}
i\partial_tG^r(t,t^{\prime})=\delta(t-t^\prime)\sum_{\bm{ij}}\langle\{\Psi_{\bm{i}}(t), \Psi_{\bm{j}}^{\dagger}(t^\prime)\}\rangle+\theta(t-t^\prime)\sum_{\bm{ij}}\langle\{[\Psi_{\bm{i}}(t), \mathcal{H}_{full}(t)], \Psi_{\bm{j}}^{\dagger}(t^\prime)\}\rangle.
\end{align}
After a double Fourier transform $\frac{1}{2\pi}\int{dt^\prime}e^{-i\epsilon^\prime t^\prime/\hbar}\int{dt}e^{i\epsilon t/\hbar}i\partial_t]G^r(t,t^{\prime})$, we then have
\begin{align}
\epsilon G_C^r(\epsilon,\epsilon^{\prime})=\delta(\epsilon-\epsilon^\prime)+\{\sum_\alpha\mathcal{H}_\alpha+\mathcal{H}_T+H_0+\sum_{k\bm{i}}[T_x\mathcal{J}_k(y\Phi_0)+T_x^\dagger\mathcal{J}_k(-y\Phi_0)]\}G^r(\epsilon+k\hbar\omega,\epsilon^{\prime}).
\label{free_C_green_2e}
\end{align}
It is obvious that $G^r(\epsilon,\epsilon^{\prime})$ is nonzero only when $\epsilon-\epsilon^{\prime}=n\hbar\omega$ with $n$ being an integer. Eq.~\ref{free_C_green_2e} is thus equivalent to
\begin{align}
(\epsilon+m\hbar\omega) G_{mn}^r(\epsilon)=\delta_{mn
}+\{\sum_\alpha\mathcal{H}_\alpha+\mathcal{H}_T+H_0+\sum_{k\bm{i}}[T_x\mathcal{J}_k(y\Phi_0)+T_x^\dagger\mathcal{J}_k(-y\Phi_0)]\}G_{(m+k)n}^r(\epsilon).
\label{eom}
\end{align}
If $\Phi_0=0$ or $\omega=0$ (see Eqs.~\ref{Lattic_Hamiltonian_phase} and ~\ref{Full_Hamiltonian}), the system returns to the equilibrium case and the Green's function is only a function of time difference. When $\Phi_0=0$, $\mathcal{J}_k(0)=\delta_k$, in this case Eq.~\ref{eom} can be simplified into $(\epsilon+m\hbar\omega) G_{mn}^r(\epsilon)=\delta_{mn
}+\{\sum_\alpha\mathcal{H}_\alpha+\mathcal{H}_T+H_0+\sum_{\bm{i}}[T_x+T_x^\dagger]\}G_{mn}^r(\epsilon)$ or $\epsilon G^r(\epsilon)=I+\mathcal{H}_{full}G^r(\epsilon)$. Similarly when $\omega=0$, $\sum_k\mathcal{J}_k(x\ne0)=1$. Since the Green's functions now are only a function of time difference, Eq.~\ref{eom} can be written as $\epsilon G^r(\epsilon)=I+\{\sum_\alpha\mathcal{H}_\alpha+\mathcal{H}_T+H_0+\sum_{k\bm{i}}[T_x\mathcal{J}_k(y\Phi_0)+T_x^\dagger\mathcal{J}_k(-y\Phi_0)]\}G^r(\epsilon)$, which is also $\epsilon G^r(\epsilon)=I+\mathcal{H}_{full}G^r(\epsilon)$. Generally, the retarted Green's function $G_{mn}^r(\epsilon)$ can then be solved iteratively or self-consistently~\cite{Li2018}. In our numerical calculation, we use the wide band approximation by assuming that the Green's function of the metallic leads does not dependent on the energy.

Alternatively, the retarted Green's function can also be obtained from the Dyson equation. We separate the full Hamiltonian into two part $\mathcal{H}_{full}(t)=\mathcal{H}_0+\mathcal{H}_1(t)$ where $\mathcal{H}_0=\sum_{\alpha}\mathcal{H}_\alpha+\mathcal{H}_T+H_0$ while $\mathcal{H}_1(t)=\sum_{k\bm{i}}[T_x\mathcal{J}_{k}(y\Phi_0)+T_x^\dagger\mathcal{J}_{k}(-y\Phi_0)]\exp{(ik\omega t)}$. It is clear that $\mathcal{H}_0$ is time independent and solvable. The free Green's function on the basis of $\mathcal{H}_0$ can then be defined as
 \begin{align}
g^r(t,t^{\prime})=-i\theta(t-t^\prime)\sum_{\bm{ij}}\langle\{\Psi_{\bm{i}}(t), \Psi_{\bm{j}}^{\dagger}(t^\prime)\}\rangle_0,
\end{align}
where $\langle\cdots\rangle_0$ refers to the state average on the basis of $\mathcal{H}_0$ instead of the full Hamiltonian. According to the definition,  the Dyson equation of the Green's function is
\begin{align} 
G^r(t,t^{\prime})=g^r(t,t^{\prime})+\int{dt_1}g^r(t,t_1)\sum_{k\bm{i}}[T_x\mathcal{J}_{k}(y\Phi_0)+T_x^\dagger\mathcal{J}_{k}(-y\Phi_0)]\exp{(ik\omega t)}G^r(t_1,t^{\prime}).
\end{align}
Because $\mathcal{H}_0$ is time independent, its free Green's function $g^r(t,t^{\prime})$ is only a function of the time difference $t-t^\prime$. Therefore, we obtain the Dyson equation after a Fourier transform
\begin{align} 
G^r(\epsilon,\epsilon^\prime)=g^r(\epsilon)\delta(\epsilon-\epsilon^\prime)+g^r(\epsilon)\sum_{k\bm{i}}[T_x\mathcal{J}_k(y\Phi_0)+T_x^\dagger\mathcal{J}_k(-y\Phi_0)]G^r(\epsilon+k\hbar\omega,\epsilon^\prime),
\end{align}
or
\begin{align} 
G^r_{mn}(\epsilon)=g^r_{mm}(\epsilon)\delta(m-n)+g^r_{mm}(\epsilon)\sum_{k\bm{i}}[T_x\mathcal{J}_k(y\Phi_0)+T_x^\dagger\mathcal{J}_k(-y\Phi_0)]G^r_{(m+k)n}(\epsilon).
\label{Dyson_eq}
\end{align}
On the other hand, the free Green's function $g^r_{mm}(\epsilon)=1/[\epsilon+m\hbar\omega-\mathcal{H}_0]$, Eq.~\ref{Dyson_eq} is exactly the same with Eq.~\ref{eom} after multiplying $\epsilon+m\hbar\omega-\mathcal{H}_0$ on both sides.

The lesser Green's function $G^<_{mn}(\epsilon)$ can then be obtained using the Keldysh equation 
\begin{align}
G^<_{mn}(\epsilon)=\sum_kG^r_{mk}(\epsilon)f_k(\epsilon)[g^{a-1}_k(\epsilon)-g^{r-1}_k(\epsilon)]G^a_{kn}(\epsilon),
\end{align} 
where the Fermi distribution function $f_k(\epsilon)=1$ when $\epsilon\le 0$ while $f_k(\epsilon)=0$ otherwise.  $G^a_{mn}(\epsilon)=[G^r_{nm}(\epsilon)]^\dagger$ is the advanced Green's function.
Once we obtain the Green's function $G^<_{mn}(\epsilon)$, the current can then be calculated straightforwardly from Eq.~\ref{current_final}.